\documentstyle[12pt,epsfig]{article}
\begin{document}

\begin{center}
{\Large \bf A THREE-DIMENSIONAL CODE FOR MUON}\\
\vspace{0.2cm}
{\Large \bf PROPAGATION THROUGH THE ROCK:}\\
\vspace{0.2cm}
{\Large \bf MUSIC}

\vspace {1cm}
P.Antonioli$^{1}$, C.Ghetti$^{1}$, E.V.Korolkova$^{2}$, V.A.Kudryavtsev$^{2}$,
G. Sartorelli$^{1}$\\
\vspace{1cm}
$^{1}${\it University of Bologna and INFN-Bologna, Bologna, Italy}\\
$\,^{2}${\it Institute for Nuclear Research of the Russian Academy of Science,
Moscow, Russia}\\

\vspace{1cm}
To be published in {\it Astroparticle Physics}
\end{center}
\vspace{1cm}

\centerline{\Large \bf Abstract}

\vspace{0.5cm}

We present a new three-dimensional Monte-Carlo code MUSIC 
(MUon SImulation Code) for muon propagation through the rock.
All processes of muon interaction with matter with high energy loss
(including the knock-on electron production) are treated as
stochastic processes. The angular deviation and lateral displacement
of muons due to multiple scattering, as well as bremsstrahlung, pair
production and inelastic scattering are taken into account.
The code has been applied to obtain the energy distribution and
angular and lateral deviations of single muons at different depths
underground. The muon multiplicity distributions obtained with MUSIC
and CORSIKA (Extensive Air Shower simulation code) are also
presented. We discuss the systematic uncertainties of the results
due to different muon bremsstrahlung cross-sections.\\

\vspace{1cm}

\noindent PACS number(s): 94.40T

\pagebreak

\section{ Introduction }

Muon transport through the rock plays an important role in 'underground
physics', in particular in the underground cosmic-ray experiments. The muon
'depth-intensity curve' related to the muon energy spectrum at sea level
and to the primary cosmic-ray specrum, muon multiplicity distribution and
'decoherence curve' related to the mass composition of primaries and to
the characteristics of high-energy hadron-nucleus interactions are
influenced by the muon propagation through the rock. The large underground
detectors designed to search for rare events (such as proton decay,
neutrino interactions etc.) are not free from the background of cosmic-ray
muons and muon-produced secondary particles. The propagation of muons
produced by high-energy astrophysical and atmosheric neutrinos in the rock
should be also taken into account. These examples show the importance of
an accurate three-dimensional simulations of the muon transport in matter.

There are several three-dimensional codes for the simulation of muon
propagation through the rock (see, for example, \cite{Bilokon,Lipari,MACRO}). 
They have been used to obtain
the characteristics of muon flux and muon bundles deep underground 
and to describe the experimental data. However, the increase of
the accuracy of the recent and future measurements requests an
adequate increase of the accuracy of the calculations. The
accuracy of the calculations is restricted by the uncertainties in
the cross-sections and the assumptions made to simplify the 
computational procedure (due to the restricted CPU time).
Recently, a new corrections to the muon bremsstrahlung and knock-on
electron production cross-sections were calculated \cite{KKP}.
A new generation of powerful computers allows to avoid many
simplifications in the calculation procedure.\\

In our code, named MUSIC (MUon SImulation Code) 
we have used the most recent and accurate cross-sections
of the muon interactions with matter and we tried to avoid many
simplifications used in the previous simulations. This three-dimensional
code is based on an algorithm of a one-dimensional muon propagation 
code described in \cite{Kudryavtsev} and used to fit the 
'depth-intensity relation' measured by LVD \cite{LVD,LVDR}.
We have treated all processes of muon interaction with high energy
transfer (bremsstrahlung, inelastic scattering, pair production
and knock-on electron production) as the stochastic ones if
the fraction of energy lost by muon is more than $10^{-3}$.
The mean muon path between two interactions with the energy loss
more than $10^{-3}$ is about 20 hg/cm$^2$. The angular deviation 
of muons was taken into account not only in the
process of multiple scattering (as it was usually done) but also 
in the processes of bremsstrahlung, inelastic scattering and
pair production. We investigated the effect due to the last three
processes.\\

We have performed three-dimensional simulations of the muon transport
thro\-ugh Gran Sasso and standard rocks. The results of such simulations
for Gran Sasso rock, as well as the complete description of the code
and related topics can be found in \cite{Ghetti}. Here we present the
results of the calculations for standard rock using two different
bremsstrahlung cross-sections \cite{KKP,BBb}. In this way we 
investigated the effect due to the uncertainties in the cross-sections.
We have also applied the code to calculate muon multiplicity
distribution using CORSIKA \cite{COR} code for the simulations
of Extensive Air Showers.\\

The main features of the MUSIC code are described in Section 2. The results
of the simulation in the standard rock are presented in Section 3.
We also discuss in Section 3 the dependence of our results on the
cross-sections and assumptions used in the code.
The results of the full Monte-Carlo simulation of muon component 
using CORSIKA and MUSIC can be found in Section 4. In Section 5 we
present our conclusions.

\section{The code }

In the code we have treated energy losses due to ionization,
bremsstrahlung, pair production and 
muon -- nucleus inelastic scattering.\ 

The mean energy loss is usually expressed by

\begin{equation}
- <\frac{dE}{dx}>=\alpha (E) +\beta(E) E\
\end{equation}

\noindent where $\alpha$ is the energy loss due to ionization and $\beta$ is
the relative energy loss due to other processes.\

Ionization can be considered
as a continuous process and it is well described by
Bethe-Bloch formula. On the contrary,
in other processes a particle can lose a big fraction of its energy
in a single interaction and 
it is necessary to treat these kinds of energy losses in two separate parts,
one continuous and the other stochastic.
To improve the simulation procedure and to study the importance of
the stochasticity of the knock-on electron production we have divided
the energy loss due to this process into two parts: continuous loss
with $v<v_{cut}=10^{-3}$ which we calculate using the Bethe-Bloch
formula, and stochastic loss which we treat in the same way as other
stochastic processes. Thus, the energy losses due to knock-on electron
production, bremsstrahlung, pair production and inelastic scattering
are expressed by:\

\begin{eqnarray}
<\frac{dE}{dx}>
& = &
<\frac{dE}{dx}>_{cont}+<\frac{dE}{dx}>_{stoc} \nonumber\\
& = & E \frac{N}{A}\int_{0}^{v_{cut}} dv v \frac{d\sigma}{dv} +  
E \frac{N}{A}\int_{v_{cut}}^{1} dv v \frac{d\sigma}{dv}\
\end{eqnarray}

\noindent where A is the mass number of the material,\ N is 
the Avogadro number,\
$v$ is the fraction of the transferred muon energy and
$\frac{d\sigma}{dv}$ is the cross-section of the process. \

The differential probability that a muon loses a fraction $v$ of its energy
per g/cm$^{2}$ in a stochastic way is given by\

\begin{equation}
\frac{dP}{dv} =\frac{N}{A}\frac{d\sigma}{dv}
\end{equation}

It is quite difficult to choose a suitable value for $v_{cut}$,\ because
a little value of $v_{cut}$, that means an accurate simulation,
results in an increase of CPU's time. We have chosen  $v_{cut} = 10^{-3}$
that seems meet all requirements. On a DEC AlphaStation 250 4/266 
the simulation of 10 TeV muons propagated through 3 km.w.e. takes 0.02 s/muon.

We note that most of the other codes use higher value of 
$v<v_{cut}$ and/or
are considering the knock-on electron production as a continuous
process.
  
The aim of the code is to propagate a muon  through
a layer X of standard rock (we used A=22, Z=11, $\rho$= 2.65 g/cm$^{3}$).\ 
If we know the initial energy of the particle
$E_{0}$ , its starting point $x_{0},\ y_{0}, z_{0}$
in the rock and its direction $\theta_{0},\ \phi_{0}$,\
we can simulate with an iterative procedure,\ step by step,\  muon 
interaction point
according to $P(|\overrightarrow{r}|)=\frac{1}{\lambda}
e^{-|\overrightarrow{r}|/\lambda(\sigma)}$,
where $\overrightarrow{r}$ 
is the vector which defines the interaction point in the
space and $\lambda$ is the mean free path. Then the continuous 
and stochastic energy losses, as well as
lateral and angular displacement due to
multiple Coulomb scattering are evaluated. Moreover the code
simulates the angular deviations caused by the 
stochastic processes.\

Finally we check if the muon has crossed the layer X,\ and in
this case we store
its final energy $E_{f}$ ,\ its direction $\theta_{f},
\ \phi_{f}$
and the coordinates of its
arrival point  $x_{f},\ y_{f}, z_{f}$.\

To describe pair production we have used the cross-section given by 
Kokoulin-Petrukhin \cite{pairKP}. The code samples separately muon 
energy loss and angular deflection. 
The second one is sampled following the parametrization proposed in
\cite{VanGinneken} (see Appendix A).

Two different muon bremsstrahlung cross-sections,\ one given 
by Bezrukov and Bugaev \cite{BBb} and the other
recently proposed by Kelner, Kokoulin and Petrukhin \cite{KKP} are
available in MUSIC.
These two calculations differ mainly by the account of bremsstrahlung
on atomic electrons, when the photon is emitted by the target electron.

The code samples the fraction of muon energy transferred to the photon 
and separately the muon scattering angle, following also in this
case the parametrisation proposed by \cite{VanGinneken}
(see Appendix A). 

We carried out additional test of the simulation of muon deflection
angle due to pair production and bremsstrahlung process using a simplified
procedure based on the GEANT parametrizations \cite{pairGEANT,bgeant}.
We concluded that this approach leads to a
little underestimation of the muon deflection due to these processes with
respect to Van Ginneken's parametrizations.

%We verified that the approach proposed by GEANT \cite{pairGEANT,bgeant}
%for muon deflection due to pair production and bremsstrahlung brings
%to a little underestimation of the effect with respect to Van 
%Ginneken parametrisation.

Among stochastic processes, the angular deflection due to inelastic scattering
on nuclei is expected to be dominant \cite{giapuICRC}.
In this case to calculate the muon energy loss and angular deflection  
we have used double-differential cross section 
$\frac{d^{2}\sigma}
{dv d Q^{2}}$ obtained by Bezrukov and Bugaev \cite{nuclBB}, 
where Q is the four-momentum transferred.
We have sampled the muon scattering angle $\theta_{\mu}$ according 
to the aforementioned cross-section
using  the relation  $Q^{2}= 2(E_{\mu 0}E_{\mu f} - P_{\mu 0}P_{\mu f}cos
\theta_{\mu}- m_{\mu}^{2})$. Since the muon inelastic scattering is
expected to be the most important for angular deviation of muons (with the
exception of multiple scattering), this process has been treated in a
more correct way (using the double-differential cross-section) 
in comparison with other stochastic processes. However, the
cross-section of this process proposed in \cite{nuclBB} is valid
only at low $Q^2$ (the accuracy is not worse than $11\%$ for 
$Q^2 \leq 3$ GeV$^2$ and energy transfers from 1 to 10$^6$ GeV
\cite{nuclBB}).\

Some authors \cite{VanGinneken,MoeTsai,Battistoni}, 
pointed out the complexity of the problem to sample simultaneously
the emission angle of the photon, the deflection angle of the muon 
and the fractional energy loss $v$. The accurate and simultaneous
simulation of the angles and energies can be important for muons exceeding
1 TeV, where the deflection due to bremsstrahlung could be dominant
\cite{VanGinneken}.
As it will be discussed in the next section, however, even if
we used the same approach as in \cite{VanGinneken}, we didn't find
such behaviour, and the multiple
scattering appears to be the dominant mechanism of deflection of muons
passing through the large thickness of matter.
A complete treatment of these aspects is currently
beyond the aim of our code.

We have treated multiple Coulomb scattering in the Gaussian approximation
\cite{Rossi}. We have considered the projections of muon angular deviation, 
$\theta_{\mu}$, and lateral
displacement, $l$, on two orthogonal planes which include the initial
direction of muon. The projections of angular deviation and lateral
displacement of muon with respect to its initial direction
at each step between two
interactions with $v>v_{cut}$
are distributed according to correlated Gaussians with
mean values and dispersions calculated using the formulae:\

\begin{equation}
<\theta_{x}> = <\theta_{y}> = 0, \ \ \  
\sigma_{\theta}^{2} = <\theta_{x}^{2}> = <\theta_{y}^{2}> = 
\frac{\epsilon _{ms}^{2}
}{2\lambda}  \int_{0}^{x}
\frac{dx}{E_{\mu}^{2}}\
\end{equation}

\begin{equation}
<l_x> = <l_y> = 0 \ \ \   
\sigma_l^2 = <l_x^{2}> = <l_y^{2}> = \frac 
{\epsilon _{ms}^{2}}{6\lambda \rho^{2}}
\int_{0}^{x} \frac{dx x^{2}}{E_{\mu}^{2}}\
\end{equation}
 
\noindent where $\epsilon _{ms}$ = 21.2 MeV, $\lambda$ = 26.48 g/cm$^{2}$
is the radiation length in standard rock and
$\rho$ = 2.65 $g / cm^{3}$ is the standard rock density.
As discussed in the next section, we have checked also if the
use of a more accurate treatment of multiple scattering
(e.g. Moli\`ere theory) leads to different results.

\section{Results}

We have propagated 100000 muons through 3 km.w.e. and 
1000000 through 10 km. w.e. with MUSIC using Kelner-Kokoulin-Petrukhin's  
cross-section for bremsstra\-hlung.
Initial muon energy has been sampled according to a primary energy spectrum
with index $\gamma=3.7$.\ In the case of 3 km.w.e. minimal muon energy
$E_{0}$ was fixed to 900 GeV,\ while for 10 km.w.e. $E_{0}$ was 9 TeV,
corresponding in each case to a survival probability of less or about 0.3\%.

\begin{figure}[ht]
\begin{center}
\mbox{\epsfig{figure=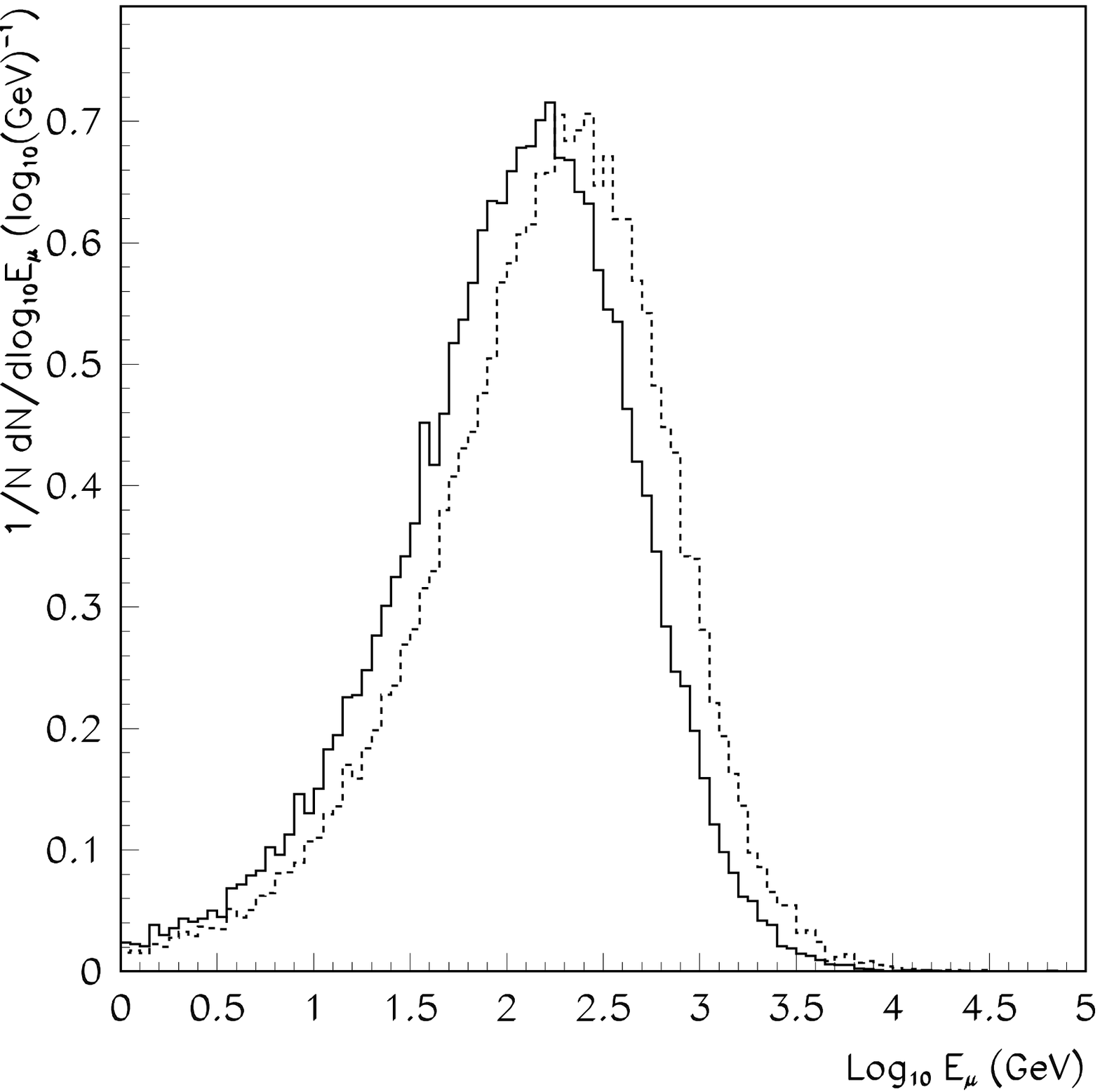,width=15.cm,height=17.cm}}
\end{center} 
\caption{Energy distributions of muons (from a power spectrum
with index $\gamma$ = 3.7) at 3 km.w.e. (solid line) and at 
10 km.w.e. (dashed line).}
\label{fi:fig1}
\end{figure}

\begin{figure}[ht]
\begin{center}
\mbox{\epsfig{figure=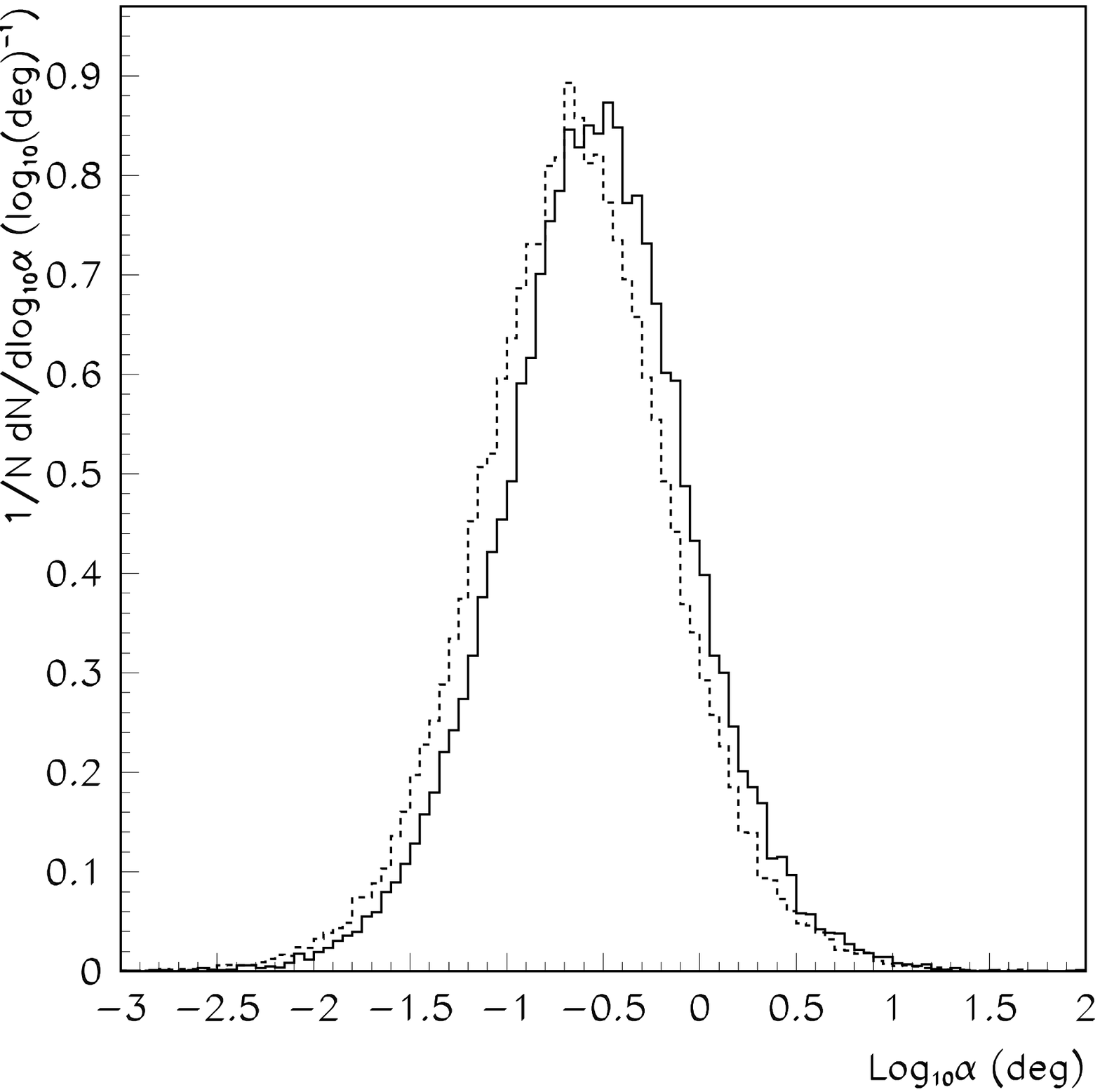,width=15.cm,height=17.cm}}
\end{center} 
\caption{Angular distributions of muons (from a power spectrum
with index $\gamma$ = 3.7) at 3 km.w.e. (solid line) and at 
10 km.w.e. (dashed line).}
\label{fi:fig2}
\end{figure}

\begin{figure}[ht]
\begin{center}
\mbox{\epsfig{figure=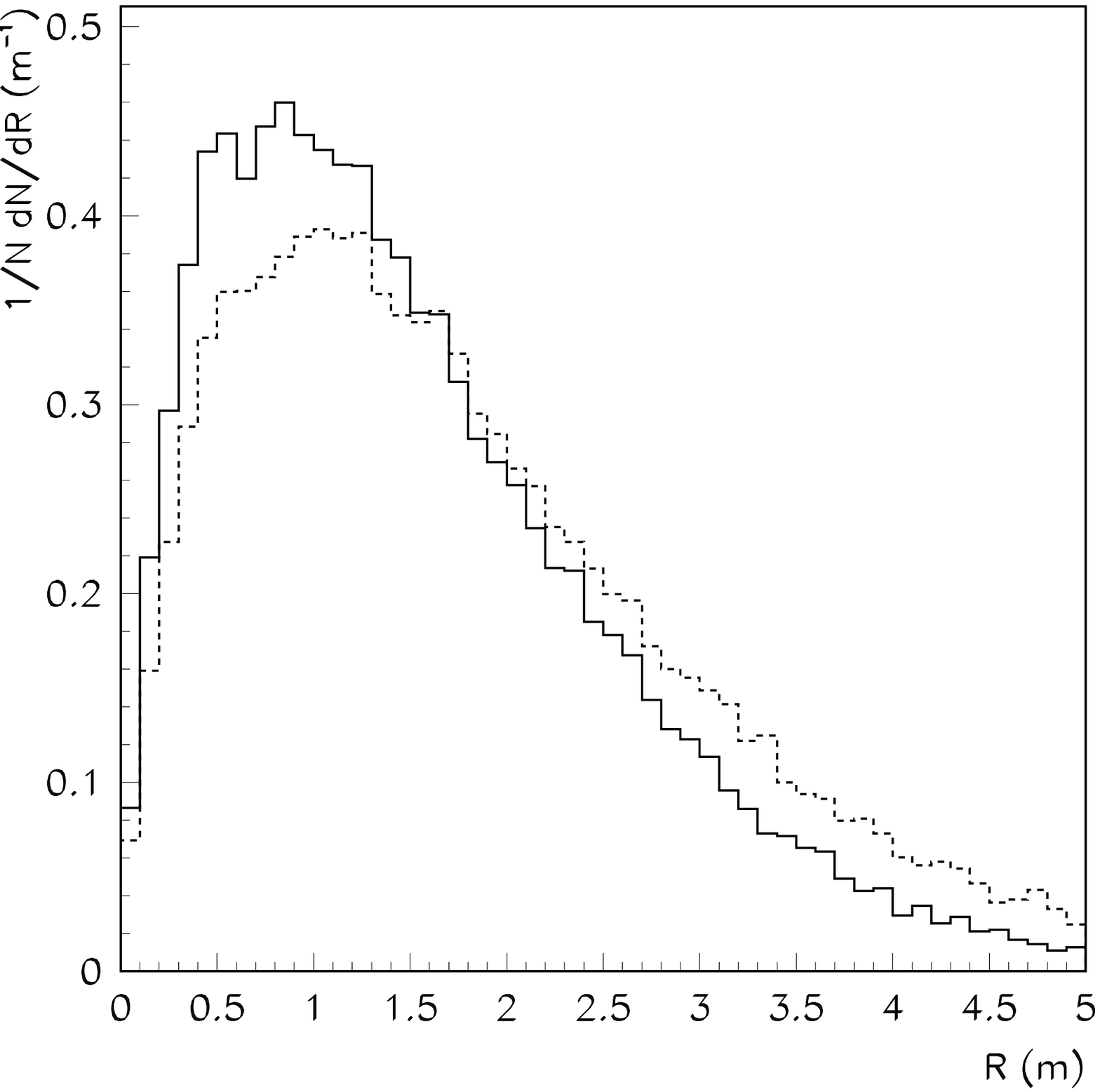,width=15.cm,height=17.cm}}
\end{center} 
\caption{Lateral distributions of muons (from a power spectrum
with index $\gamma$ = 3.7) at 3 km.w.e. (solid line) and at 
10 km.w.e. (dashed line).}
\label{fi:fig3}
\end{figure}

Figures \ref{fi:fig1}, \ref{fi:fig2} and \ref{fi:fig3} show energy, lateral 
and angular distributions
at 3 and 10 km.w.e.\ The mean muon energy 
is 250 GeV
at 3 km.w.e.\ and 367  GeV at 10 km.w.e. 
This result is in agreement with the simple consideration that
the mean muon energy at a depth $h$  increases with $h$.

Mean angular deviation $<\alpha>$
is  0.56$^{\circ}$ at 3 km.w.e.\ and 0.45$^{\circ}$ at 10 km.w.e.\ Angular 
deviation is mainly
caused by multiple Coulomb scattering.\ This  process produces
a deviation that increases when energy decreases.\ This explains why at 3
km.w.e.,\ when $<E_{h}>$ is lower,\ $<\alpha>$ is higher than at 10 km.w.e.\

\begin{figure}[ht]
\begin{center}
\mbox{\epsfig{figure=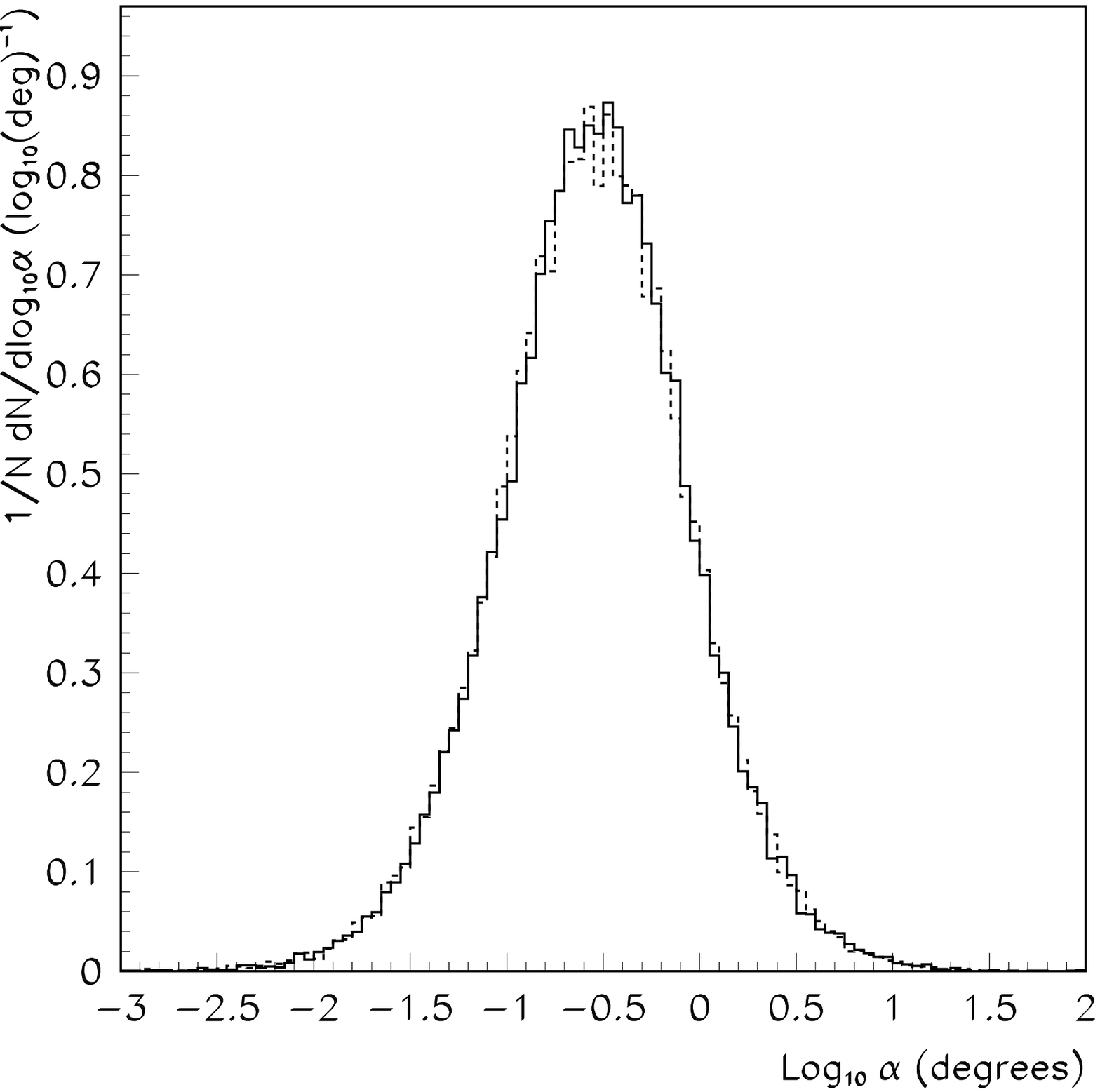,width=15.cm,height=17.cm}}
\end{center} 
\caption{Comparison of angular distributions 
obtained with MUSIC (solid line) and PROPMU (dashed line)
at 3 km.w.e.}
\label{fi:fig4}
\end{figure}

\begin{figure}[ht]
\begin{center}
\mbox{\epsfig{figure=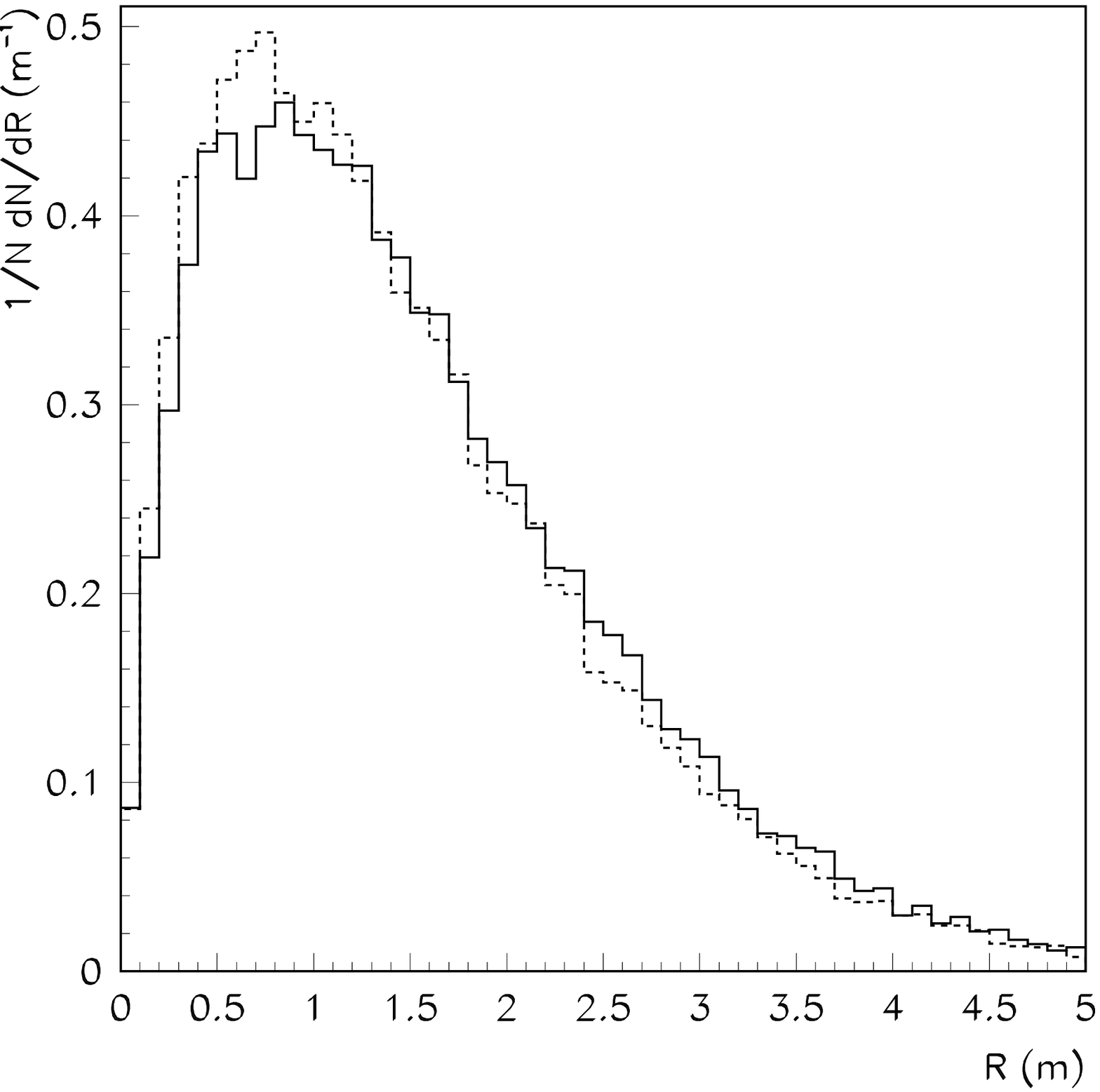,width=15.cm,height=17.cm}}
\end{center} 
\caption{Comparison of lateral distributions 
obtained with MUSIC (solid line) and PROPMU (dashed line)
at 3 km.w.e.}
\label{fi:fig5}
\end{figure}

Figures \ref{fi:fig4} and \ref{fi:fig5} show lateral 
and angular distributions obtained with MUSIC and
PROPMU -- a code for muon propagation in rock written by P. Lipari
and T. Stanev \cite{Lipari}. The codes are completely independent,\ but 
their results are
practically the same.\ This fact confirms the correctness of both 
simulations and shows that the effects of angular deviations due to
stochastic processes, with the approximations discussed in previous
section, are really very little, 1-2 orders of magnitude less than 
the effects due to multiple scattering.
In fact, as Lipari and Stanev treat only
angular deviation due to multiple scattering, and in MUSIC all possible angular
deviations are taken into account,\ the distributions obtained with these two
codes do not show any appreciable difference.\

\begin{figure}[ht]
\begin{center}
\mbox{\epsfig{figure=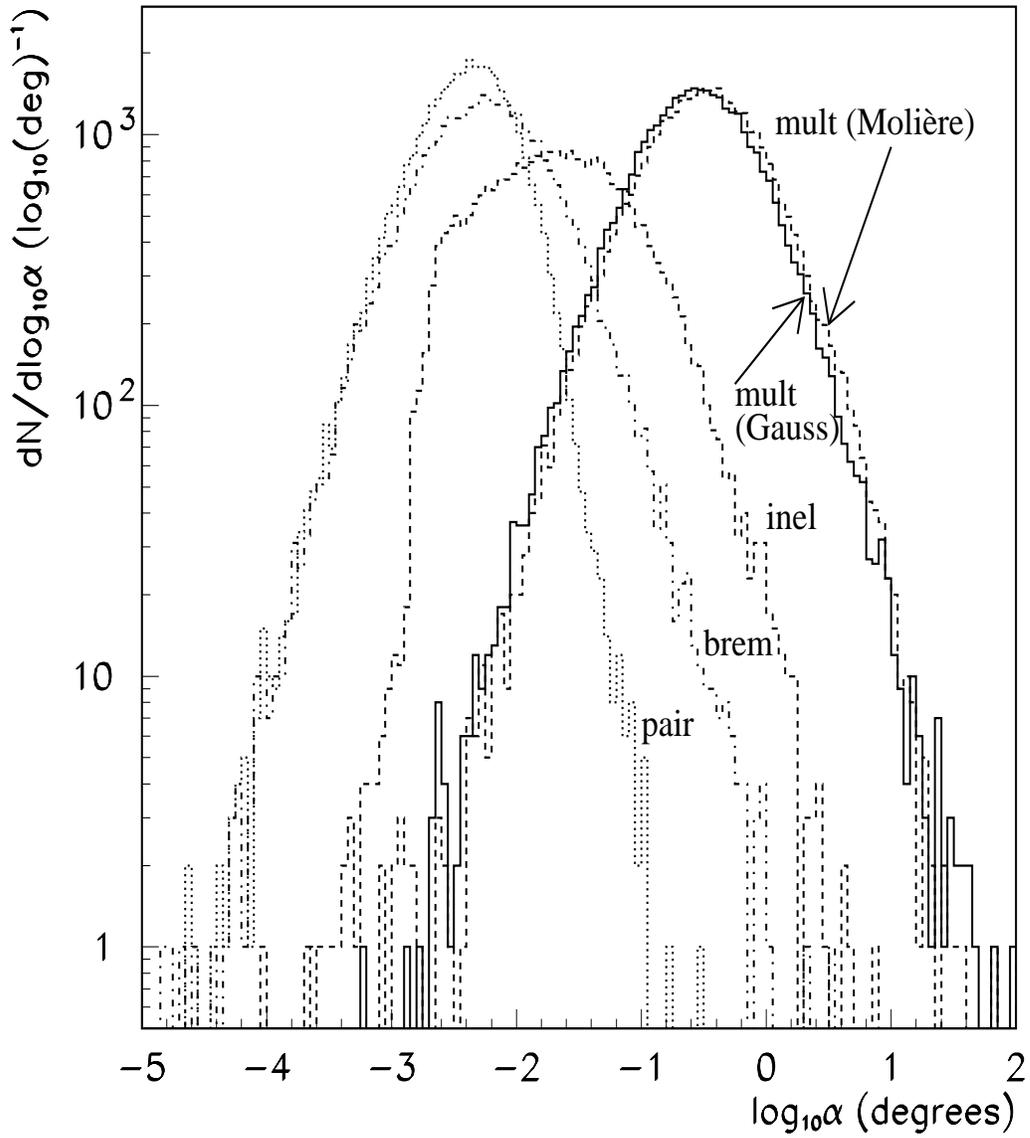,width=15.cm,height=17.cm}}
\end{center} 
\caption{Angular deviations generated by each mechanism of
deflection separately at 3 km.w.e. (from a power spectrum
with index $\gamma$ = 3.7): multiple scattering (continuous: Gaussian
theory, dashed: Moli\`ere theory),
inelasting scattering (dashed), bremsstrahlung (dot-dashed),
pair production (dotted).}
\label{fi:fig6}
\end{figure}

In Figure \ref{fi:fig6} we show the contribution of each process to the
angular deviation of muons which pass 3 km.w.e. To do this plot
we have repeated 
%four 
five times the same simulation leaving active 
one mechanism of deflection each time. For multiple
scattering we have used either Gaussian treatment
described in previous section or Moli\`ere theory
subroutines provided by GEANT \cite{GEANT_Mol}.
A more detailed treatment (Moli\`ere theory)
results in the similar distribution compared with that of Gaussian
treatment, with a small increase of mean deflection
angle.
However, the original Moli\`ere theory 
(see \cite{GEANT_Mol} and references therein) does not provide
the lateral displacement, more
important from the experimental point of view.
As can be clearly seen from Figure \ref{fi:fig6} the multiple
scattering dominates over other processes. 
This result remains true for 10 km.w.e.
and for a monochromatic beam of 10 TeV muons.

\begin{figure}[ht]
\begin{center}
\mbox{\epsfig{figure=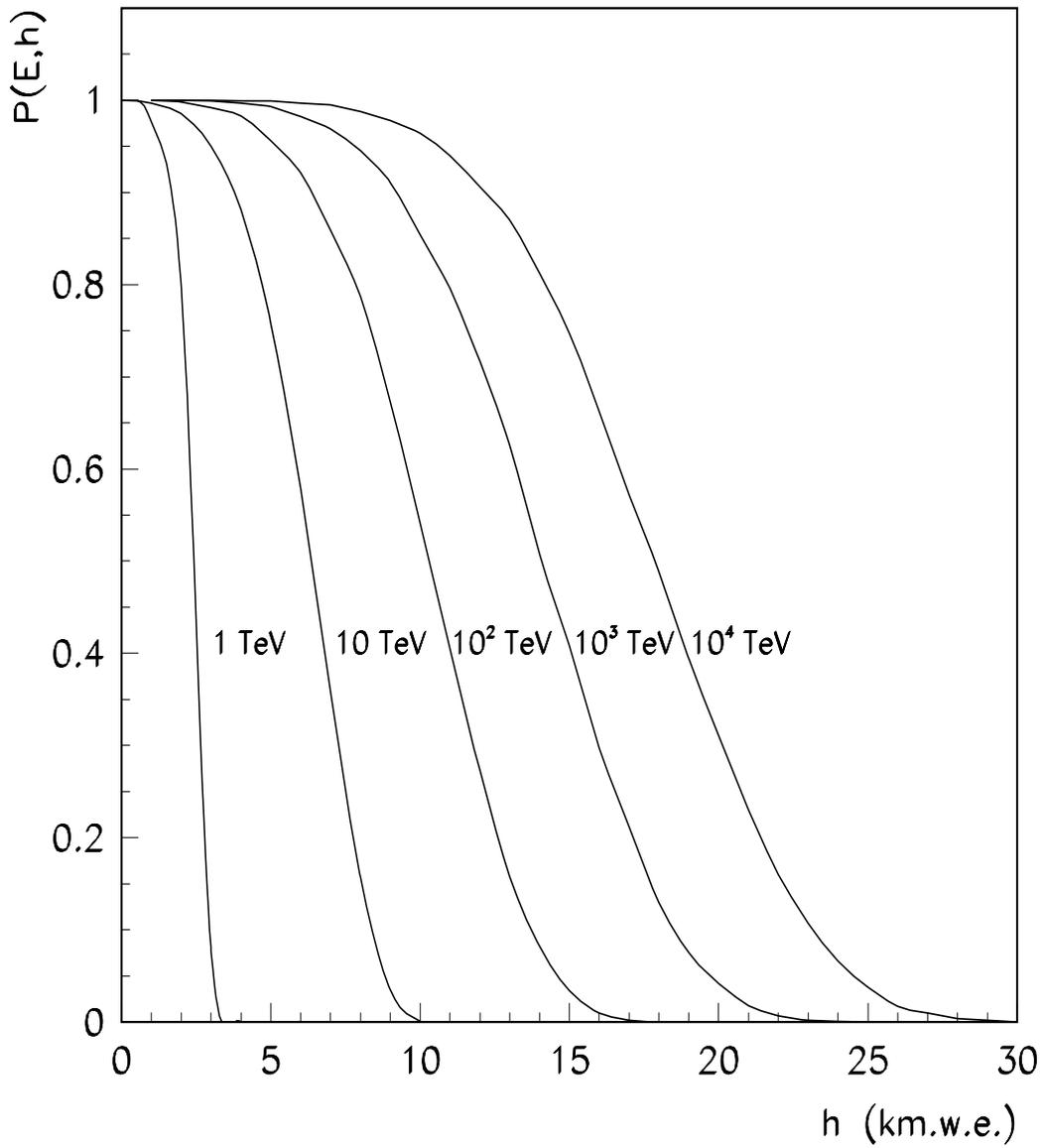,width=15.cm,height=17.cm}}
\end{center} 
\caption{Survival probabilities obtained with MUSIC
for $E_{0}$ from 1 to 10$^{4}$ TeV.}
\label{fi:fig7}
\end{figure}

\begin{figure}[ht]
\begin{center}
\mbox{\epsfig{figure=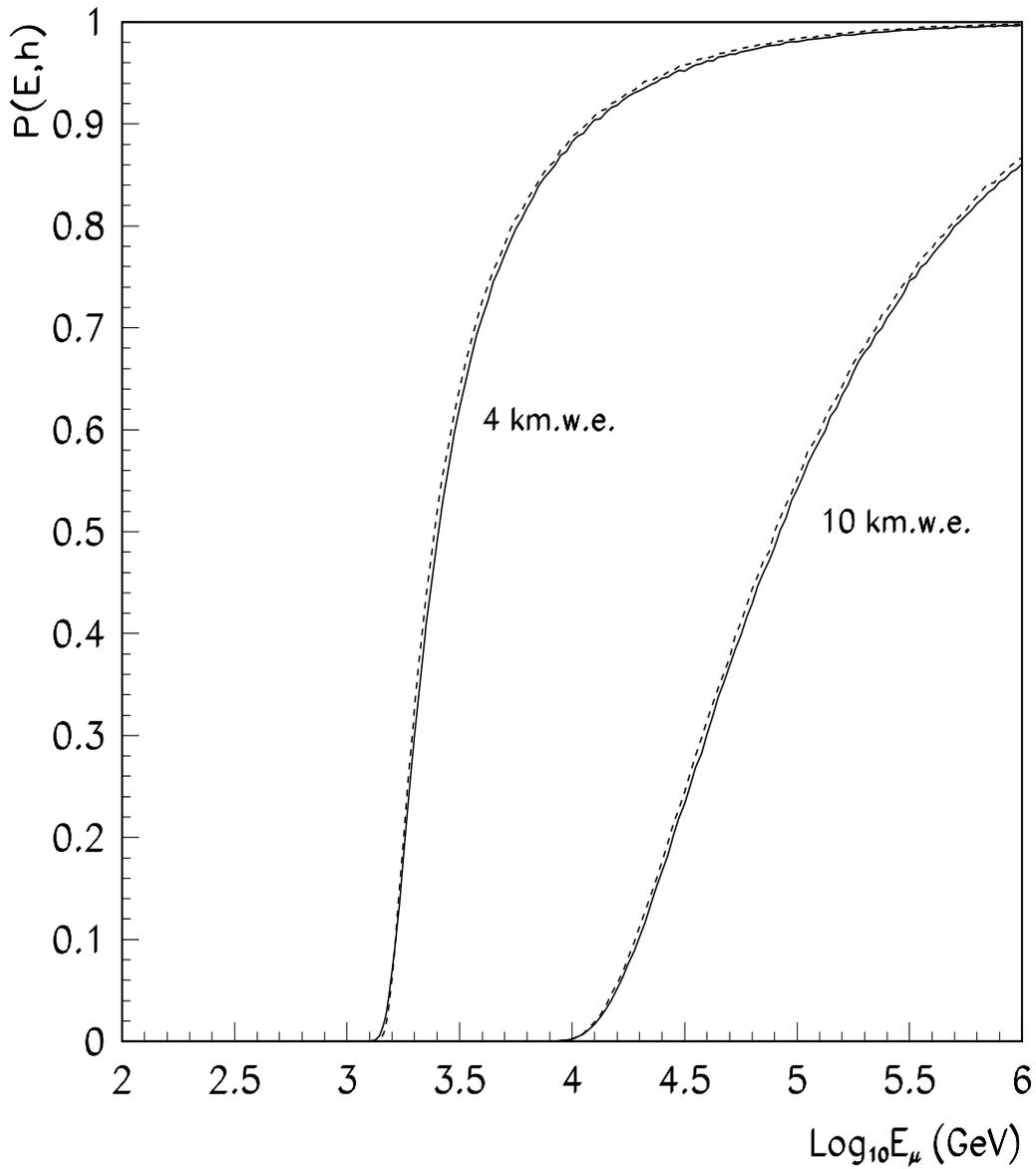,width=15.cm,height=17.cm}}
\end{center} 
\caption{Comparison between survival probabilities 
obtained with MUSIC (solid line) and PROPMU (dashed line)
at 4 km.w.e. and 10 km.w.e.}
\label{fi:fig8}
\end{figure}

To calculate survival probabilities, $P (E_{0},h )$,
and muon energy distributions underground, $P (E_{0},h,E)$, 
we have propagated muons
with different initial energy (100000 muons for each initial energy)
through the  standard rock. 
$P (E_{0},h,E)$ means the probability that muon with initial energy
$E_{0}$ will reach the depth $h$ with energy $E$.
Figure \ref{fi:fig7} shows 
the survival propabilities obtained
with MUSIC for $E_{0}$ from 1 to $10^{4}$ TeV.\ Figure \ref{fi:fig8}
shows $P(E_{0},h)$
for h = 4 km.w.e. and h = 10 km.w.e., calculated with MUSIC and with PROPMU.
The curves are very similar also in this case. Our values are on average 
(1-2)\% lower: this difference can be due to the new cross-section 
for bremsstrahlung used in MUSIC.\

We have studied the effects of the cross-section difference and of the
stochasticity of different processes on the characteristics of the muon
flux underground. Using the sea-level vertical muon energy spectrum,
proposed by T.Gaisser \cite{Gaisser} with the spectral index of
primary spectrum $\gamma=2.70$, and the distributions
$P (E_{0},h)$ and $P (E_{0},h,E)$ we have calculated the vertical
muon intensities $I(h)$ and mean muon energies $<E(h)>$ at different
depths underground:

\begin{equation}
I(h)=\int_0^{\infty}P(E_{0},h) \cdot {{d I(E_{0})}
\over
{d E_{0}}} \cdot d E_{0},          \label{intensity}
\end{equation}

\begin{equation}
<E(h)>={{\int_0^{\infty} E \cdot
{{d I(E,h)} \over {d E}} \cdot d E} \over
{\int_0^{\infty}{{d I(E,h)} \over {d E}} \cdot
d E}} 					\label{mean energy}
\end{equation}

\noindent where ${{dI(E_0)}\over{dE_0}}$ is the vertical muon energy spectrum 
at sea level and ${{dI(E,h)}\over{dE}}$ is the vertical muon spectrum
at the depth $h$. The latter one is obtained using the formula:

\begin{equation}
{{d I(E,h)} \over {d E}}=\int_0^{\infty}
P(E_{0},h,E) \cdot
{{d I(E_{0})}\over
{d E_{0}}} \cdot d E_{0},          \label{spectrum}
\end{equation}

We have calculated the muon intensities and mean muon energies for two
different muon bremsstrahlung cross-sections (from \cite{KKP} and 
\cite{BBb}) and for two values of $v_{cut}$ ($10^{-3}$ and $10^{-2}$).
We have carried out additional tests considering the knock-on
electron production and pair production as continuous processes.
The results of the calculations are presented in Table \ref{ta:intmu} for
$I(h)$ and Table \ref{ta:enemu} for $<E(h)>$. The first column in both tables
shows the depth in km.w.e. of standard rock. Other columns show the intensity
in units $cm^{-2} s^{-1} sr^{-1}$ (Table \ref{ta:intmu}) and the mean muon energy
in $GeV$ (Table \ref{ta:enemu}). 2nd column present our basic values calculated
with the muon bremsstrahlung cross-section from \cite{KKP} and 
$v_{cut}=10^{-3}$. The results obtained with bremsstrahlung cross-section
from \cite{BBb} and $v_{cut}=10^{-3}$ are presented in column 3.
The values in column 4 were calculated with the cross-section from \cite{KKP}, 
$v_{cut}=10^{-3}$ and with continuous energy loss due
to knock-on electron production. The simulation related to column 5
has been carried out under the same assumptions as that for column 4
but both knock-on electron and pair production have been treated as
continuous processes. Finally, column 6 shows the results of the
simulation similar to that for column 4, but with $v_{cut}=10^{-2}$.
The accuracy of the values presented in Tables \ref{ta:intmu} and 
\ref{ta:enemu} is of the order of $1\%$.

\begin{table}[htb]
\caption{
Vertical muon intensities, $I(h)$, in units
$cm^{-2} s^{-1} sr^{-1}$ for different
depths in the standard rock:
column 1 -- depth in km.w.e.;
column 2 -- bremsstrahlung cross-section from [4],
$v_{cut}=10^{-3}$, all processes are stochastic;
column 3 -- bremsstrahlung cross-section from [9],
$v_{cut}=10^{-3}$, all processes are stochastic;
column 4 -- bremsstrahlung cross-section from [4],
$v_{cut}=10^{-3}$, continuous energy loss due to
knock-on electron production ($e$ -- cont.);
column 5 -- bremsstrahlung cross-section from [4],
$v_{cut}=10^{-3}$, continuous energy loss due to
knock-on electron production and pair production ($e+p$ -- cont.);
column 6 -- bremsstrahlung cross-section from [4],
$v_{cut}=10^{-2}$, continuous energy loss due to
knock-on electron production.
}

\begin{center}
\begin{tabular}{|c|c|c|c|c|c|}
\hline
$h$    & \cite{KKP}  & \cite{BBb} & \cite{KKP} & \cite{KKP} & \cite{KKP}\\
km.w.e.&$v_{cut}=10^{-3}$&$v_{cut}=10^{-3}$&$v_{cut}=10^{-3}$&
$v_{cut}=10^{-3}$&$v_{cut}=10^{-2}$\\
       &             &            &$e$--cont. &$e+p$--cont.&$e$--cont.\\
\hline
1  &$1.23\cdot10^{-6}$ &$1.28\cdot10^{-6}$ &$1.22\cdot10^{-6}$ &
$1.22\cdot10^{-6}$ &$1.22\cdot10^{-6}$\\
2  &$1.23\cdot10^{-7}$ &$1.30\cdot10^{-7}$ &$1.21\cdot10^{-7}$ &
$1.21\cdot10^{-7}$ &$1.21\cdot10^{-7}$\\
3  &$2.37\cdot10^{-8}$ &$2.53\cdot10^{-8}$ &$2.33\cdot10^{-8}$ &
$2.34\cdot10^{-8}$ &$2.33\cdot10^{-8}$\\
4  &$5.92\cdot10^{-9}$ &$6.33\cdot10^{-9}$ &$5.79\cdot10^{-9}$ &
$5.78\cdot10^{-9}$ &$5.78\cdot10^{-9}$\\
5  &$1.69\cdot10^{-9}$ &$1.81\cdot10^{-9}$ &$1.65\cdot10^{-9}$ &
$1.64\cdot10^{-9}$ &$1.64\cdot10^{-9}$\\
6  &$5.19\cdot10^{-10}$&$5.57\cdot10^{-10}$&$5.04\cdot10^{-10}$&
$5.00\cdot10^{-10}$&$5.02\cdot10^{-10}$\\
7  &$1.66\cdot10^{-10}$&$1.80\cdot10^{-10}$&$1.62\cdot10^{-10}$&
$1.59\cdot10^{-10}$&$1.62\cdot10^{-10}$\\
8  &$5.53\cdot10^{-11}$&$5.94\cdot10^{-11}$&$5.36\cdot10^{-11}$&
$5.23\cdot10^{-11}$&$5.35\cdot10^{-11}$\\
9  &$1.87\cdot10^{-11}$&$2.02\cdot10^{-11}$&$1.82\cdot10^{-11}$&
$1.75\cdot10^{-11}$&$1.81\cdot10^{-11}$\\
10 &$6.35\cdot10^{-12}$&$6.86\cdot10^{-12}$&$6.19\cdot10^{-12}$&
$5.91\cdot10^{-12}$&$6.15\cdot10^{-12}$\\
11 &$2.19\cdot10^{-12}$&$2.38\cdot10^{-12}$&$2.15\cdot10^{-12}$&
$2.01\cdot10^{-12}$&$2.11\cdot10^{-12}$\\
12 &$7.58\cdot10^{-13}$&$8.21\cdot10^{-13}$&$7.43\cdot10^{-13}$&
$6.86\cdot10^{-13}$&$7.24\cdot10^{-13}$\\
\hline
\end{tabular}
\end{center}
\label{ta:intmu}
\end{table}

\begin{table}[htb]
\caption { Mean muon energies at vertical, $<E(h)>$, 
in $GeV$ for different
depths in the standard rock (see caption of Table 1).}

\begin{center}
\begin{tabular}{|c|c|c|c|c|c|}
\hline
$h$    & \cite{KKP}  & \cite{BBb} & \cite{KKP} & \cite{KKP} & \cite{KKP}\\
km.w.e.&$v_{cut}=10^{-3}$&$v_{cut}=10^{-3}$&$v_{cut}=10^{-3}$&
$v_{cut}=10^{-3}$&$v_{cut}=10^{-2}$\\
       &             &            &$e$--cont. &$e+p$--cont.&$e$--cont.\\
\hline
1  & 125 & 124 & 123 & 121 & 123\\
2  & 205 & 203 & 203 & 202 & 202\\
3  & 259 & 255 & 257 & 253 & 257\\
4  & 296 & 291 & 294 & 290 & 293\\
5  & 321 & 315 & 319 & 313 & 318\\
6  & 337 & 332 & 337 & 329 & 336\\
7  & 349 & 342 & 348 & 341 & 346\\
8  & 356 & 351 & 356 & 347 & 353\\
9  & 360 & 353 & 360 & 351 & 358\\
10 & 364 & 359 & 364 & 353 & 360\\
11 & 364 & 358 & 364 & 354 & 362\\
12 & 364 & 359 & 364 & 354 & 363\\
\hline
\end{tabular}
\end{center}
\label{ta:enemu}
\end{table}

Let us first compare the intensities obtained with different
assumptions used in the code.
The intensities calculated with the muon bremsstrahlung cross-section
from \cite{BBb} are systematically higher (by (4-8)$\%$) than those with the
cross-section from \cite{KKP}. The relative difference slightly 
increases with 
depth. If the depth--intensity relation measured in some
experiment is used to obtain the power index of the muon energy
spectrum at sea level (or the primary spectrum), then,
the use of the cross-section from \cite{BBb} will result in
a higher power index (softer muon spectrum) than that with
cross-section \cite{KKP}. The difference can be of the order of
0.01.\

The treatment of the knock-on electron production and pair production
as continuous processes results in the decrease of muon intensities
(up to $10\%$ at high depths).
The relative difference also increases with depth. The power index
of the primary spectrum which can be evaluated from the underground
measurements will be lower in this case. The difference 
in the power index is
negligible if only knock-on electron production is treated as
a continuous process but can be of the
order of 0.02 if both knock-on electron and pair production are
not treated stochastically.\

The increase of $v_{cut}$ from $10^{-3}$ up to $10^{-2}$ results in 
a small decrease of the muon intensities ($(2-3)\%$). However,
the increase of the relative difference with depth is small and
may not influence much the power index which can be evaluated from
the measured depth -- intensity relation. The increase of power
index in this case should not exceed 0.01.\

All aforementioned treatments lead to the conclusion that
the reasonable cross-sections and assumptions used in the code
(may be with the exception of the continuity of the pair
production process)
result in the muon intensities which do not differ much from each
other. The difference is comparable with the systematic uncertainty
which can be attributed to the existing experimental data.
The difference in the power index of primary spectrum which can be
evaluated from the experimental data using calculated intensities
should be less than the errors (including systematical error) of
the existing data. However, the stochasticity of
pair production is quite important for this purpose.\

We would like to note that these considerations are attributed to
the single muons. The situation can change for multiple muons when
the small decrease of the muon survival probability can result
in the significant decrease of the intensity of muon bundles with
large multiplicity. 
In section 4 we present some investigations of this point.\

The mean muon energies calculated with our basic assumptions (2nd
column of Table \ref{ta:enemu}) are higher than those obtained with another
cross-section and/or other assumptions. The difference is not
negligible in the cases of the cross-section from \cite{BBb}
and/or the continuity of the pair production process.\

We have carried out also the simulation of muon transport with
our basic assumptions and with $v_{cut}=3 \cdot 10^{-4}$.
We observed no statistically significant difference in the
muon intensities and mean muon energies comparing with those
obtained with $v_{cut}=10^{-3}$. We can conclude that our
basic assumptions are optimal from the point of view of
the correctness of the results and CPU time.\

We have checked also the influence of the cross-section difference
and $v_{cut}$ on the angular and lateral distributions. The initial
sea--level muon spectrum in the power-law form with power index 3.7
has been used. With the cross-section from \cite{KKP} we have obtained
the mean angular deviation 0.56$^{\circ}$ and 0.45$^{\circ}$ for 3 and 
10 km.w.e., 
respectively, while
with the cross-section from \cite{BBb} the values are 
0.56$^{\circ}$ and 0.46$^{\circ}$.
Statistical uncertainty is of the order of 0.01$^{\circ}$. 
The mean lateral displacement
is 1.56 m and 1.87 m with the cross-section from \cite{KKP} for 3 and 10
km.w.e., respectively, and 1.59 m and 1.93 m with the cross-section
from \cite{BBb}. Statistical uncertainty is about 0.01 m. 
We can conclude that 
the difference in the cross-sections does not influence much the angular
and lateral distributions at any depth. A small but statistically
significant difference of
the mean lateral displacements at large depth can be attributed to
the difference of the mean muon energies at this depth (see Table
\ref{ta:enemu} and Eq. (5)).\

We have carried out also the propagation of muons with $v_{cut}=10^{-2}$.
Actually, the increase of $v_{cut}$ from $10^{-3}$ to $10^{-2}$ results
in an increase of the average step between two consecutive interactions
by one order of magnitude. Thus, this procedure allows us to test the
dependence of the resulting angular deviation and lateral displacement
on the step between two consecutive interactions. We didn't find any
significant difference between the results with $v_{cut}=10^{-3}$ and
$v_{cut}=10^{-2}$. This proves the correctness of the three-dimensional
simulation procedure.

\section{Muon multiplicity}

We have roughly investigated also the effect of 
bremsstrahlung cross-section on muon
multiplicity distribution. We have used the code CORSIKA \cite{COR}
to simulate the development of a shower produced by a primary cosmic-ray 
nucleus in atmosphere. We have sampled two sets of 10000 showers 
produced by protons and iron nuclei of 10$^{4}$ TeV to have a
good statistics of muons.\

\begin{figure}[ht]
\begin{center}
\mbox{\epsfig{figure=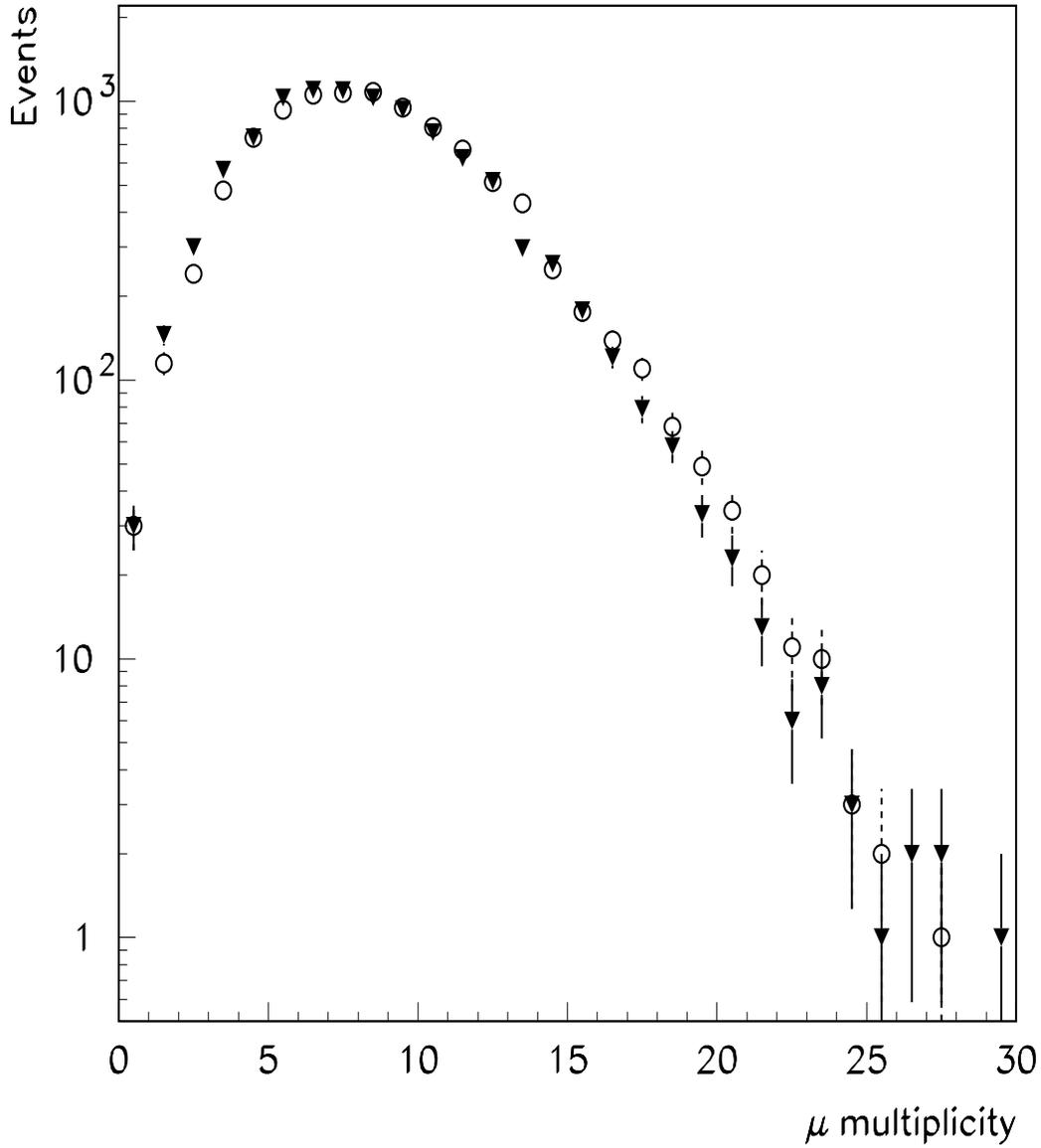,width=15.cm,height=17.cm}}
\end{center} 
\caption{Comparison of muon multiplicity distributions 
at 3 km.w.e. obtained using bremsstrahlung cross sections from [9] 
(open circles)
and [4] (triangles). Muons are produced in 10000
proton showers of energy $E_o$ $=$ 10$^{4}$ TeV}
\label{fi:fig9}
\end{figure}

\begin{figure}[ht]
\begin{center}
\mbox{\epsfig{figure=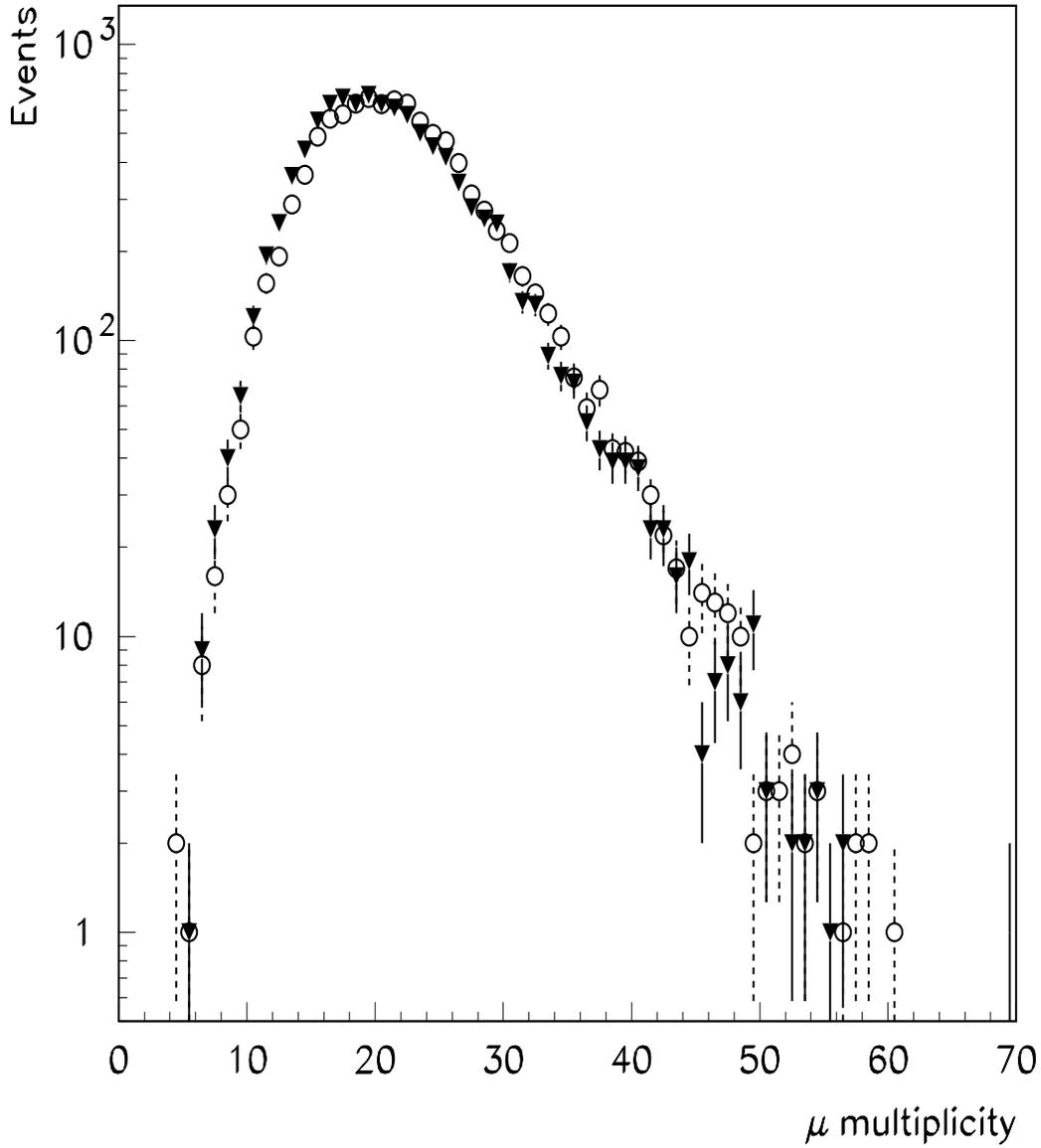,width=15.cm,height=17.cm}}
\end{center}
\caption{Comparison of muon multiplicity distributions
at 3 km.w.e obtained using bremsstrahlung cross sections from [9]
(open circles)
and [4] (triangles). Muons are produced in 10000
showers initiated by iron nuclei of energy $E_o$ $=$ 10$^{4}$ TeV}
\label{fi:fig10}
\end{figure}

To transport shower muon component through 3000 m.w.e. of 
standard rock with MUSIC, we have used two cross sections for bremsstrahlung
from \cite{BBb} and \cite{KKP}.
For the case of infinite underground detector, we have produced
two muon multiplicity distributions obtained with different cross-sections.
Figures \ref{fi:fig9} and \ref{fi:fig10} show these distributions 
for showers produced by protons and
iron nuclei. In Table \ref{ta:mulmu} the mean values $<N_{\mu}>$ are
presented, their percentage difference 
$\frac{<N_{\mu BB}> - <N_{\mu KKP}>}{<N_{\mu KKP}>}$ is $\sim 4\%$.\

\begin{table}[hb]
\caption{Mean muon multiplicity at 3 km.w.e. for proton and iron
primaries of energy $10^4$ TeV obtained 
using different bremsstrahlung cross-sections.}
\begin{center}
\begin{tabular}{|c|c|c|}
 \hline
$Cross-section$ 
& $<N_{\mu}>\ p  $ & $<N_{\mu}>\ Fe  $  \\ \hline
$ \cite{KKP} $ 
& $7.89 \pm 0.04  $ & $ 20.96 \pm 0.07 $ \\ \hline
$ \cite{BBb}  $ & $ 8.20 \pm 0.04  $ &$ 21.71 \pm 0.07$
 \\ \hline
\end{tabular}
\end{center}
\label{ta:mulmu}
\end{table}  

In order to evaluate the possibility that the differences between distributions
could be caused by statistical fluctuactions, we have done a $\chi ^{2}$
test: the probability that muon multiplicity distributions obtained with
different cross sections come from a single distribution is $\sim 10\%$.\

\begin{figure}[ht]
\begin{center}
\mbox{\epsfig{figure=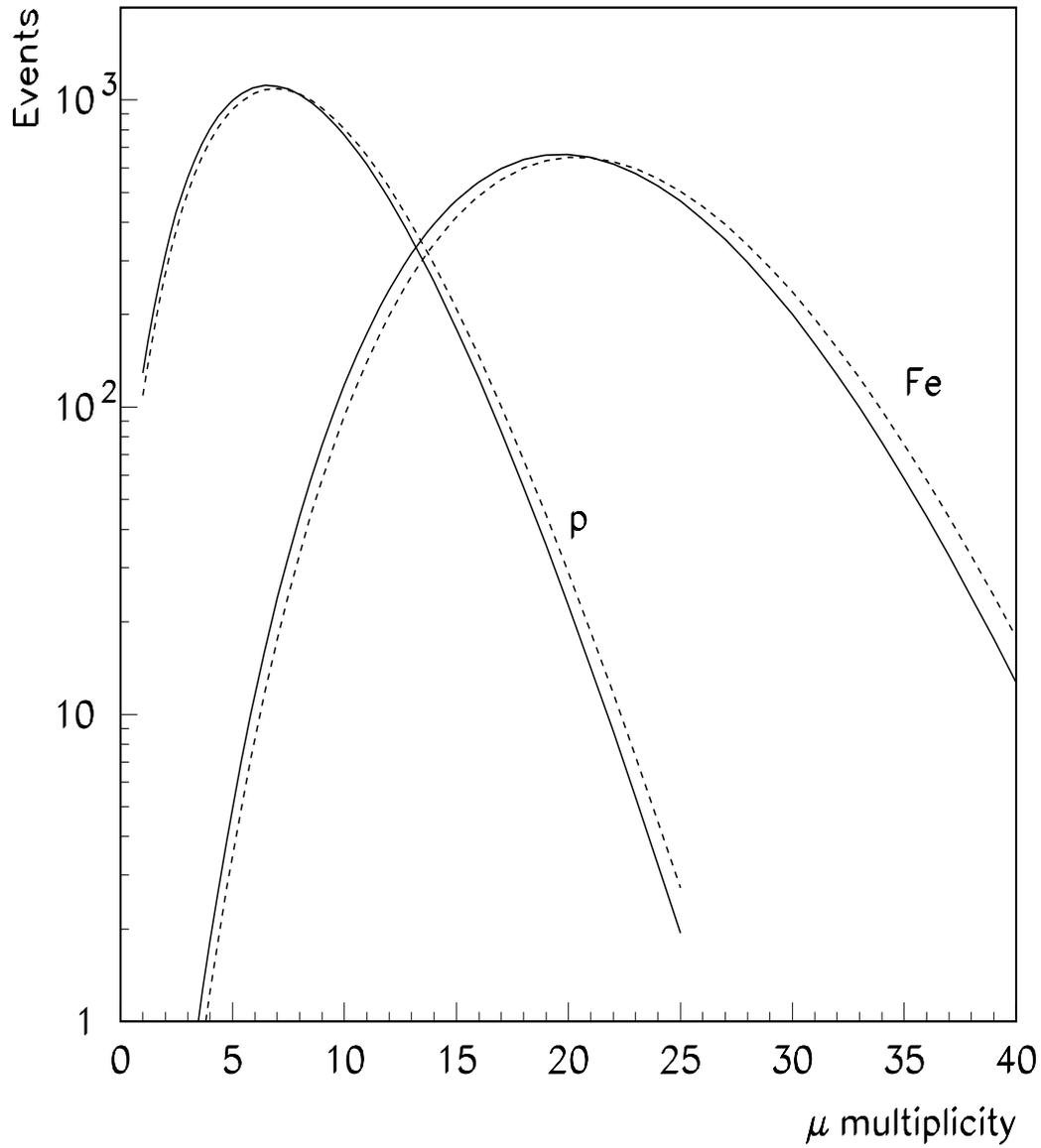,width=15.cm,height=17.cm}}
\end{center} 
\caption{Comparison of the fits to the multiplicity distributions with
negative binomial functions. Multiplicity distributions have been
obtained using muon bremsstrahlung cross sections from [9] (dashed line)
and [4] (solid line) for proton and iron primaries.}
\label{fi:fig11}
\end{figure}

The effect of different bremsstrahlung cross sections on muon 
multiplicity can be underlined using a fit with a negative binomial
\cite{Forti}.  It is evident, as shown in Fig. \ref{fi:fig11}, 
that fit to muon multiplicity distribution, obtained with
cross-section from \cite{BBb}, 
is practically a translation of the other fit due to 
different values of $<N_{\mu}>$.\ 

These simple considerations proves that the muon interaction
cross-sections can influence the muon multiplicity distributions
underground and, hence, the primary cosmic-ray composition which
can be evaluated from the underground measurements.\

In real experiments, however, the observed multiplicity distribution
is driven by the primary energy spectrum and the assumed 
primary composition and a such complete simulation to investigate the
differences due to the muon cross sections is beyond the aim of this paper. 
We can state only that the effect is quite small (but not negligible)
and dominated by other systematic uncertainties.

\section{Conclusions}

We have presented the new three-dimensional Monte-Carlo code
MUSIC for muon propagation through the rock. The main advantages
of this code are: i) the treatment of all main
processes of muon interaction with matter
(bremss\-trah\-lung, inelastic scattering, pair production
and knock-on electron production)
with the fractional
energy loss $v > 10^{-3}$ as stochastic processes;
ii) the use of the most accurate cross-sections of muon interactions
with matter;
iii) the accounting of
the angular deviation of muons
due to stochastic processes.
We can conclude that the angular and lateral distributions
of muons at any depth are fully determined by multiple
Coulomb scattering, while other processes give a contribution
which is 1-2 orders of magnitudes less. However, the very rare processes
of muon scattering to large angles (for example, backward
scattering) which can be due to inelastic scattering and/or bremsstrahlung
need more accurate treatment of these processes at large $Q^2$.
We have studied the dependence of the characteristics of muon flux
underground on the bremsstrahlung cross-section and assumptions
used in the code. We can conclude that
the treatment of pair production as a stochastic process is quite
important for the characteristics of muon flux underground.
The separation of stochastic and continuous energy losses
at $v_{cut}=10^{-3}$ is optimal from the point of view of CPU time
and the accuracy of the results. 
The accurate knowledge of the cross-sections is
needed to calculate the intensity of single muons and muon bundles
underground. Even a small (1-2)\%
increase of the muon bremsstrahlung cross-section \cite{KKP}
comparing with that from \cite{BBb} results
in a significant (6\%) decrease of the single muon intensity at 3 km.w.e.
and in a shift (however, quite small) 
of muon multiplicity distributions toward low multiplicities.
This could produce some bias in the interpretation of the muon
'depth--intensity relation' and multiplicity distribution
measured in an underground experiment when they are analysed
in terms of the primary spectrum and composition.

\section{Acknowledgements}

We are grateful to Profs. G.Navarra, O.G.Ryazhskaya, and Drs. P.Lipari,
R.P.Ko\-ko\-ulin and G.Battistoni for useful discussions.
This work is supported by 
Italian Institute for Nuclear Physics (INFN),
Italian Ministry of University and Scientific-Technological
Research (MURST),
Russian State Committee of Science and Technologies
and Russian Fund of Fundamental Researches (grant 96-02-19007).

\vspace{1.0cm}

\noindent {\Large \bf Appendix A. Simulation of muon angular deviation due to
stochastic processes}

\vspace{0.5cm}

To sample the deflection angle of muon due to bremsstrahlung or pair 
production processes we followed the approximations for mean angle
$<\theta^2>^{1/2}$ between the incident and scattered muon proposed in 
\cite{VanGinneken}. 

Let us consider first the muon bremsstrahlung process.
Since for both energy loss and angular deviation
the so-called 'coherent nuclear bremsstrahlung' 
\cite{MoeTsai,VanGinneken} is the most important, we have
used the parametrisation for $<\theta^2>^{1/2}$ of \cite{VanGinneken} for 
this process:

\begin{eqnarray}
<\theta^2>^{1/2} 
& = &
\left[ \left( k_1v^{1/2},k_2 \right) _{min}, k_3v \right] _{max},
v \leq 0.5 \nonumber \\
& = &
k_4 v^{1+n} (1-v)^{-n}, <\theta^2>^{1/2}<0.2, v>0.5 \\
& = &
k_5(1-v)^{-1/2}, <\theta^2>^{1/2} \geq 0.2, v>0.5 \nonumber
\label{eq:brems}
\end{eqnarray}

\noindent where $k_1=0.092E^{-1/3}$, $k_2=0.052 E^{-1} Z^{-1/4}$, $k_3=
0.22 E^{-0.92}$, $k_4=0.26 E^{-0.91}$, $n=0.81 E^{1/2}/(E^{1/2}+1.8)$,
and $k_5$ is related to $k_4$ by continuity of the function.

The mean angle between incident and scattered muon due to pair
production process has been parametrised as \cite{VanGinneken}:

\begin{eqnarray}
<\theta^2>^{1/2}
& = &
(2.3+\ln E) E^{-1} (1-v)^{-1} (v-2m_e/E)^2 v^{-2} \nonumber \\
& \times &
\left[ 8.9 \cdot 10^{-4} v^{1/4} (1+1.5 \cdot 10^{-5} E)+
0.032v/(v+1), 0.1 \right] _{min}
\label{eq:pair}
\end{eqnarray}

\noindent where $m_e$ is the electron mass.

For a given value of $v$ sampled according to the cross-sections
from \cite{KKP,pairKP} without any additional approximation
the mean angle $<\theta^2>^{1/2}$ between incident
and scattered muon has been calculated using the Eqs. (9,10). 
The spatial angle $\theta$ has been chosen from an
exponential in the variable $\theta^2$ with the mean value 
$<\theta^2>$.

We note again that we did not use any additional parametrisation
for the muon inelastic scattering cross-section which has been taken
from \cite{nuclBB}. For a given value of $v$ sampled according to
the cross-section integrated over $Q^2$, the muon scattering angle
$\theta$ has been simulated using the double-differential cross-section.

For all stochastic processes we have sampled randomly the angle $\phi$
in the plane perpendicular to the muon direction.

\vspace{1cm}

The code is available upon the request by e-mail to:
antonioli@bo.infn.it or kudryavtsev@vaxmw.tower.ras.ru

\pagebreak

\pagebreak

\end{document}